\documentclass[reprint,aps,prd,amsmath,amssymb,floatfix,nofootinbib,preprintnumbers]{revtex4-2}

\usepackage{graphicx}
\usepackage[dvipsnames]{xcolor}
\usepackage{hyperref}
\hypersetup{colorlinks, urlcolor=BlueViolet, citecolor=Plum, linkcolor=PineGreen}
\usepackage{bm}
\usepackage{cleveref}
\usepackage{physics}
\usepackage[utf8]{inputenc}
\usepackage{amsmath}
\usepackage{amssymb}
\usepackage{enumitem}
\usepackage{relsize}
\usepackage{blindtext}
\usepackage{tabulary}

\graphicspath{Figures/}

\begin{document}

\title{Cosmic-ray dark matter confronted by constraints on new light mediators}
\date{September 2023}

\author{Nicole F.~Bell}
\email{n.bell@unimelb.edu.au}
\affiliation{ARC Centre of Excellence for Dark Matter Particle Physics$,$ \\~School of Physics$,$~ The~ University~ of~ Melbourne$,$~ Victoria~ 3010$,$~ Australia}

\author{Jayden L.~Newstead}
\email{jnewstead@unimelb.edu.au}
\affiliation{ARC Centre of Excellence for Dark Matter Particle Physics$,$ \\~School of Physics$,$~ The~ University~ of~ Melbourne$,$~ Victoria~ 3010$,$~ Australia}

\author{Iman Shaukat Ali}
\email{ishaukatali@student.unimelb.edu.au}
\affiliation{ARC Centre of Excellence for Dark Matter Particle Physics$,$ \\~School of Physics$,$~ The~ University~ of~ Melbourne$,$~ Victoria~ 3010$,$~ Australia}

\begin{abstract}
The detectability of light dark matter in direct detection experiments is limited by the small kinetic energy of the recoiling targets. Thus, scenarios where dark matter is boosted to relativistic velocities provide a useful tactic to constrain sub-GeV dark matter particles. Of the possible dark matter boosting mechanisms, cosmic-ray upscattering is an appealing paradigm as it doesn't require any additional assumptions beyond dark matter coupling to nucleons or electrons. However, detectable signals are obtained only with relatively large cross sections which, in turn, can only be realized with large couplings, light mediators or composite dark matter. In this work we consider a general set of light mediators that couple dark matter to hadrons. Using data from Borexino, XENON1T, LZ and Super-K, we show that existing constraints on such mediators preclude appreciable cosmic-ray dark matter upscattering. This finding highlights the limited applicability of cosmic-ray upscattering constraints and suggests they only be used in model-dependent studies.
\end{abstract}

\maketitle

\section{Introduction}
The existence of dark matter (DM) is established through evidence of its gravitational influence on galactic (and larger) scales, yet its particle nature and composition remain unknown. Direct detection experiments aim to observe nuclear or electronic recoils resulting from collisions of DM particles from the galactic halo with detector targets. These experiments have placed increasingly strong constraints on heavy DM candidates. 
However, all direct detection experiments, even those with the lowest energy thresholds, rapidly lose sensitivity to sub-GeV DM, because nuclear recoils from non-relativistic DM-nucleon scattering become undetectable. The strongest constraints on the DM-nucleon cross section for low-mass DM are set by cosmology and astrophysics at $\sigma_{\chi N} \sim 10^{-28}\text{cm}^2$ \cite{Nadler:2019zrb,Boddy:2018kfv,Gluscevic:2017ywp,Slatyer:2018aqg,Xu:2018efh,Bhoonah:2018wmw,Bhoonah:2018gjb,Wadekar:2019mpc}, some $\sim15$ orders of magnitude weaker than direct detection constraints on GeV-scale DM.

Recently, there have been two key ideas aimed at improving the low-mass sensitivity of direct detection experiments: leverage the Migdal effect to detect electronic ionisation or excitation rather than a nuclear recoil~\cite{Ibe:2017yqa,Dolan:2017xbu,Bell:2019egg,Essig:2019xkx, Bell:2021zkr,Bell:2023uvf}, or {\it give the DM a boost}. Energetic DM, with momenta much larger than that carried by standard halo-DM, would permit greatly increased energy deposition in nuclear recoil events, recovering experimental sensitivity to the scattering of sub-GeV DM particles. Such an energetic flux may originate from astrophysically boosted DM, as studied in \cite{DeRocco:2019jti,Baracchini:2020owr,An:2017ojc,An:2021qdl,Granelli:2022ysi,Emken:2017hnp,Chen:2020gcl,Emken:2021lgc,Kouvaris:2015nsa,Huang:2013xfa,Yin:2018yjn}.

One scenario of growing interest is the boosting of DM by galactic cosmic rays (CRDM) \cite{Bringmannpospelov19, Ema:2018bih,capiellobeacon2019,newsteadshoemaker2020,Guo20,Dent:2020syp,bellnewstead21,Xia:2021vbz,Alvey:2019zaa,Wang:2019jtk}. If DM interacts with nucleons or electron, DM upscattering on cosmic ray nucleons or electrons will {\it unavoidably} produce a small population of relativistic DM. 
The detection of this CRDM can then proceed via scattering off electron or nucleon targets in direct detection experiments. Alternatively, CR-DM interactions may have an observable impact on the CR flux~\cite{Cappiello:2018hsu} or produce gamma rays~\cite{Hooper:2018bfw,Dent:2020qev}.

Observable signals of CRDM in direct detection experiments are controlled by {\it two} powers of the DM scattering cross section. This is because the DM must scatter twice: first in the boosting interaction, and then again in the detection process. As a result, the experimental sensitivity is restricted to relatively large values of the scattering cross section. In fact, the relevant cross sections are typically so large that bounds from unitarity and perturbativity become an important consideration~\cite{Digman:2019wdm,Alvey:2022pad}. It is therefore important to properly evaluate physical constraints on the interaction mechanism.

Contact interactions require that the scattering cross section be smaller than the geometric size of the nucleus, $\sigma \sim 10^{-27}\text{cm}^2$. However, employing a light mediator allows for long-range interactions that can exceed this limit~\cite{Digman:2019wdm}. It is important to note that the momentum transfer in CR-DM interactions can be large, necessitating the use of the full (model-dependent) cross section, especially if the interaction is mediated by a light mediator with $m_\text{med}^2\lesssim q^2$~\cite{newsteadshoemaker2020}. Fortuitously, employing the full momentum-dependent cross section provides stronger bounds for all DM of mass $m_\chi<$ 1 GeV. A similar conclusion is reached in \cite{Dent:2020syp,Bardhan:2022bdg}, where CR-electron upscattering is studied. However, given that the mediators are required to couple to nucleons or electrons quite strongly, they themselves can be directly constrained~\cite{Digman:2019wdm,Bondarenko:2019vrb,Alvey:2022pad}. New light degrees of freedom are strongly constrained by astrophysics and beam-dump experiments, restricting the usefulness of cosmic-ray upscattering for these kinds of models.

When considering large DM-matter cross sections, it is necessary to include the effects of attenuation in the overburden on the incoming DM flux~\cite{Zaharijas:2004jv}. For all underground direct detection (or neutrino) experiments this places a ceiling on the constraints that can be obtained. Previous studies have treated the attenuation of boosted DM to varying degrees of complexity. For example, including attenuation due to inelastic scattering processes can reduce the cross section ceiling by an order of magnitude~\cite{Su:2022wpj}. As we shall see, the precise cross section above which the CRDM flux is attenuated below threshold is of lesser importance given that it is already ruled out by other experiments. Therefore, we shall neglect the additional complexity of including inelastic scattering and only treat the Earth stopping effects due to elastic scattering. 

The purpose of this paper is to study a general set of light mediators and assess whether cosmic-ray upscattering of DM remains viable in the face of constraints imposed on the mediator. Using the direct detection of CRDM in both direct detection and neutrino experiments, we compute the coupling of these light mediators to sub-GeV DM, and compare our results to constraints from other observations and experiments.

\section{CRDM with light mediators}
\label{sec:models}
We describe the DM-nucleon interactions with simplified models as in \cite{newsteadshoemaker2020}, but here we consider mediators of four types: scalar, $\phi$; vector, $V_\mu$; axial, $A_\mu$; and pseudoscalar, $\eta$. The scalar and vector mediators give rise to spin-independent (SI) scattering interactions, while the axial and pseudoscalar produce spin-dependent (SD) interactions. The effective interaction Lagrangian coupling these mediators to nucleons, $N$, and a fermionic DM candidate, $\chi$, take the form:

\begin{align}
    \mathcal{L}_\text{int}^\text{scalar}
    & = g_{\chi \phi} \phi \Bar{\chi}\chi 
    +g_{N\phi} \phi \Bar{N}N, 
    \label{eq:modelS}
    \\
    \mathcal{L}_\text{int}^\text{vector}
    & =  g_{\chi V} V^{\mu} \Bar{\chi}\gamma_{\mu}\chi
    +g_{NV} V^{\mu} \Bar{N}\gamma_{\mu}N,  
    \label{eq:modelV}
    \\
    \mathcal{L}_\text{int}^\text{axialvector}
    & =  g_{\chi A} A^{\mu} \Bar{\chi}\gamma_{\mu}\gamma^{5}\chi
    +g_{NA} A^{\mu} \Bar{N}\gamma_{\mu}\gamma^{5}N,  
    \label{eq:modelAV}
    \\
    \mathcal{L}_\text{int}^\text{pseudoscalar}
    & =  g_{\chi \eta} \eta \Bar{\chi}\gamma^{5}\chi
    +g_{N \eta} \eta \Bar{N}\gamma^{5}N,
    \label{eq:modelPS}
\end{align}
where all of the $g_{xx}$ are dimensionless coupling constants. We will assume that the mediator couples equally to neutrons and protons. 

These four mediators enable DM-nucleon scattering interactions, with cross sections (in the rest frame of the target) given by:
\begin{align}\label{eq:diffCSCRDMScalar}
    \left(\frac{d\sigma_{it}}{dT_t}\right)_{\text{scalar}}
    =&\,g_{t \phi}^2 g_{i \phi}^2 A_i^2 F_S^2(q^2) \nonumber\\
    &
    \times{\frac{ (2m_t + T_t)(4m_i^2+2m_t T_t)}
    {16\pi (T_i^2+2m_i T_i)(m_\phi^2+2m_t T_t)^2}},
\end{align}

\begin{align}\label{eq:diffCSCRDMVec}
    &\left(\frac{d\sigma_{it}}{dT_t}\right)_{\text{vector}}
    =  g_{t V}^2 g_{iV}^2A_i^2F_V^2(q^2) \nonumber\\
     &\times\frac{ (2m_{t}(m_i + T_i)^2 - ((m_i + m_t)^2+2m_t T_i)T_t + m_t T_t^2) }
    {{4\pi(T_i^2+2m_{i}T_{i})(m_{V}^2+2m_{t}T_{t})^2}},
\end{align}
\begin{align}\label{eq:diffCSCRDMAV}
    &\left(\frac{d\sigma_{it}}{dT_{t}}\right)_{\text{axial}}
    =\,g_{t A}^2 g_{iA}^2F_A^2(q^2)\nonumber\\
    &
    \times{\frac{2 m_t \left(T_i^2+3m_i^2-T_i T_t+\frac{T_t^2}{2}\right)+T_t (m_i-m_t)^2}
    {{4\pi(T_{i}^2+2m_i T_i)(m_{A}^2+2m_{t}T_{t})^2}}},
\end{align}
\begin{align}\label{eq:diffCSCRDMPS}
    \left(\frac{d\sigma_{t i}}{dT_{t}}\right)_{\text{pseudoscalar}}
    =&\, g_{t \eta}^2 g_{i\eta}^2F_P^2(q^2)\nonumber\\
    \times &\frac{m_t T_t^2}{8\pi(T_{i}^2+2m_i T_i)(m_{\eta}^2+2m_{t}T_{t})^2},
\end{align}
where the $F$ are form factors and $m_\phi$, $m_V$, $m_A$, $m_\eta$ are masses of the scalar, vector, axial and pseudoscalar mediators, respectively. These cross sections match those derived in \cite{Alvey:2022pad} and \cite{Ema:2020ulo}.
The incident and target particle masses are $m_i$ and $m_t$, while $T_i$ and $T_t$ are the incoming and outgoing kinetic energies, respectively. For the boosting interaction, the $\chi$ is the target and the CR nucleon is the incident particle. For the detection interaction, the $\chi$ becomes the incident particle and the target is a nucleus in the detector.

For SI proton (helium) scattering, the form factor $F^2 (q^2)$ takes the dipole form with mass $\Lambda_\text{p} = 0.770\,\text{GeV}\,  (\Lambda_\text{He}=0.410\,\text{GeV})$~\cite{Perdrisat:2006hj}. For proton scattering via an axial current, we use the dipole form with the axial mass, $\Lambda_\text{p}=$1.026 GeV~\cite{PhysRevD.92.113011}. The pseudoscalar form factor is related to the axial form factor by $F_\text{pseudoscalar}(q^2) = F_\text{axial}(q^2) C_q/(q^2+M_q^2)$, where we take the couplings to be isoscalar, giving $C_q=0.9\,\text{GeV}^2$ and $M_q = 0.33\,\text{GeV}$~\cite{Ema:2020ulo}. For DM scattering with nuclei larger than the proton, we take the analytic approximation of the Helm form~\cite{PhysRev.104.1466}. 

\section{CRDM Framework}

\subsection{Upscattering}

\begin{figure*}[t]
    \includegraphics[width=1.0\columnwidth]{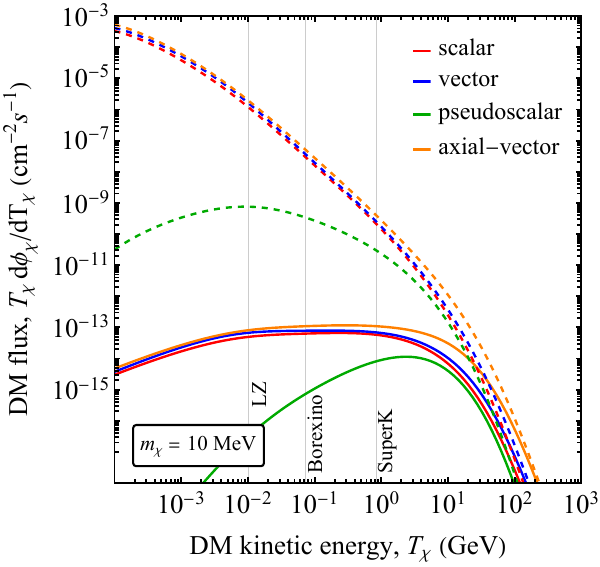}
    \hfill 
    \includegraphics[width=1.0\columnwidth]{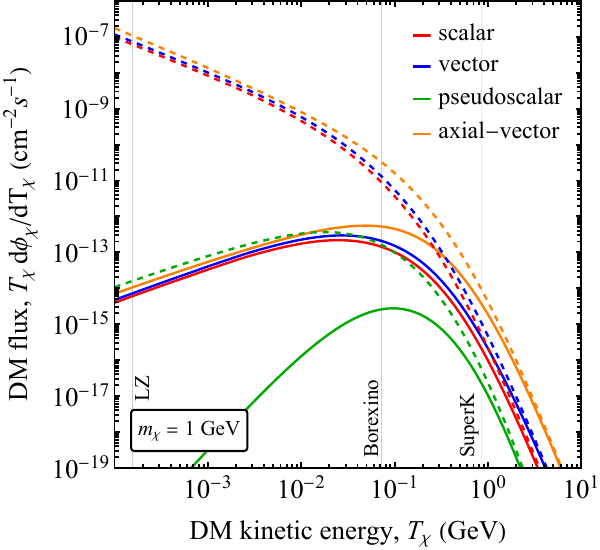}
    \caption{The CRDM flux for two example DM masses: 10 MeV \textit{(left)} and 1 GeV \textit{(right)}. The dashed and solid lines represent mediator masses of $m_\text{med}=1$ MeV and $m_\text{med}= 1$ GeV respectively. The vertical lines denote the minimum CRDM energy required to produce above-threshold recoil signals in the given experiment.}
    \label{fig:CRDM Flux}
\end{figure*}

The interactions of Eqs.~(\ref{eq:modelS})-(\ref{eq:modelPS}) enable CRs propagating in the galaxy to scatter on non-relativistic DM particles in the galactic halo. This produces an upscattered population of relativistic CRDM. Following~\cite{Bringmannpospelov19}, the differential flux at Earth is given by
\begin{align}\label{eq:fluxCRDM}
    \frac{d\Phi_{\chi}}{dT_{\chi}}
    = D_\text{eff}\frac{\rho^\text{local}_{\chi}}{m_{\chi}}
    \sum_{i}
    \int_{T^\text{min}_{i}} dT_{i} 
    \frac{d\sigma_{\chi i}}{dT_{\chi}}
    \frac{d\Phi_{i}^\text{LIS}}{dT_{i}},
\end{align}    
where $\rho_\chi^\text{local}$= 0.3 GeV $\text{cm}^{-3}$ is the local density of the DM halo and $D_\text{eff}$ is the effective diffusion zone parameter. We conservatively consider a diffusion zone of 1 kpc centered on the Earth, resulting in $D_\text{eff}=0.997$ kpc. We include contributions from cosmic ray species $i=\{p, \textrm{He}\}$ for SI scattering, and $i=p$ alone for the SD case.  The quantity ${d\Phi_{i}^\text{LIS}}/{dT_{i}}$ is then the flux of species $i$ in the local interstellar cosmic ray spectrum (LIS), we which we take from~\cite{Boschini:2017fxq}. The integration limit in the galactic rest frame is
\begin{align}\label{eq:TxMinMax}
    T_i^\text{min} =
    &\frac{T_\chi-2m_i}{2} \nonumber\\
    +
    &\frac{1}{2}\sqrt{\frac{T_\chi(2m_\chi+T_\chi)(4m_i^2+2m_\chi T_\chi)}{2m_\chi T_\chi}},
\end{align}
which is the minimum kinetic energy required by an incoming cosmic ray to upscatter DM of mass $m_\chi$ to energy $T_\chi$. 

Figure~\ref{fig:CRDM Flux} displays example DM fluxes for light and heavy DM and mediator masses. In the low momentum transfer region, if the mediator mass is much smaller than the momentum transfer, $m_\text{med}^2\ll q^2$, the CR-DM cross section is effectively enhanced by $1/q^4$ resulting in a sizable increase in flux (dashed curves). The pseudoscalar case exhibits a suppression at lower energy, which can be understood by inspection of the cross section in Eq.~\ref{eq:diffCSCRDMPS}, which exclusively depends on the transferred kinetic energy squared and thus $q^4$. This feature makes pseudoscalar-mediated DM interactions particularly hard to probe via traditional (non-relativistic) direct detection, and hence CR-upscattering has the potential to greatly improve limits on this interaction type.

\subsection{Attenuation}
\indent As most direct detection experiments are located deep underground in order to suppress the rate of background events, the scattering of DM in the atmosphere and detector overburden can significantly alter the DM flux reaching the detector. For sufficiently large scattering cross sections, the incoming DM can be decelerated to energies which will fall below experimental thresholds, giving rise to a ceiling on the interaction strength that can be probed. For a detailed study of attenuation of non-relativistic DM, see \cite{Collar:1993ss,Hasenbalg:1997hs,Zaharijas:2004jv,Kouvaris:2014lpa,Kavanagh:2016pyr,Emken:2017qmp,Bramante:2018qbc}. 

The kinetic energy loss as the DM travels a distance $x$ is given by
\begin{align}\label{eq:attenuation energy loss}
    \frac{dT_\chi}{dz}
    = -\sum_{T} n_{T} 
    \int dE_{R} \frac{d\sigma}{dE_{R}} E_{R},
\end{align} 
where the sum is over the average nuclei densities of elements in the Earth, and $E_r$ is the energy lost in each collision. This can be solved numerically to find $T_\chi^{z}(T_\chi)$, the kinetic energy at detector depth $z$. 
The nuclear density at depth $z$ is determined by a weighted average of the most abundant elements in the Earth's crust \cite{Emken:2018run}. For SI scattering, this gives an average density $n_{\mathcal{N},\text{SI}}=5.81\times 10^{-19}\,\text{nuclei}/\text{cm}^{3}$ and atomic number $A_{\text{SI}}=23.8$. Only the odd isotopes are included in SD scattering, yielding $n_{\mathcal{N},\text{SD}}=6.68\times 10^{-20}\,\text{nuclei}/\text{cm}^{3}$.
Here we perform a purely elastic treatment, but it should be noted that boosted DM carries large kinetic energies and thus the momentum transfer can be sufficiently large that inelastic scattering becomes relevant. For example, quasi-elastic scattering (QES) becomes important at momentum transfer $q^2 \gtrsim 0.1$ GeV~\cite{Alvey:2022pad, Su:2022wpj}. Considering inelastic upscattering in the production of CRDM increases the flux at high energies. This improves the cross section bounds for heavy mediators \cite{Su:2023zgr, Su:2022wpj}, however the coupling limit for light mediators remains mostly unaffected. Including the QES contribution to the total cross section also changes the attenuation ceiling, lowering it by an order of magnitude compared to the purely elastic case \cite{Su:2022wpj}. However, we shall see that the exact location of the ceiling is not of interest, given it is already excluded (and often close to perturbativity limits) in the models we  consider. Therefore, for simplicity, we shall neglect inelastic scattering.

\subsection{Scattering rate}
After attenuation in the overburden, the remaining flux of upscattered DM may scatter in a detector, resulting in nuclear recoil events with kinetic energy $E_R$. The differential event rate per unit detector mass is given by:
\begin{align}\label{eq:diffrate}
     \frac{dR}{dE_{R}}
     = \frac{1}{m_T}\epsilon(E_R)\int^{T_{\chi}^\text{max}}_{T_{\chi}^\text{min}} dT_{\chi}
     \frac{d\Phi_{\chi}}{dT_{\chi}}
     \frac{d\sigma_{\chi T}}{dE_{R}},
\end{align}
where $m_T$ is the target nuclei mass and $\epsilon(E_R)$ is an $E_R$-dependent efficiency factor. The lower bound on the energy integral can be obtained from Eq.~(\ref{eq:TxMinMax}), making the substitution $i \rightarrow \chi$, $m_\chi \rightarrow m_T$, $T\chi \rightarrow E_R$ to reverse the roles of the incident and outgoing particles. We take the upper limit on the integral to be $T_\chi^\text{max}=2$~GeV, as CRDM of higher energy make a negligible contribution to the results.
In fact, almost all of the  $m_\chi= 1$ GeV flux falls below $T_\chi=2$ GeV. 

\section{Results and discussion}

DM direct detection experiments typically express their results as limits on the DM-nucleon scattering cross section. However, this approach is problematic for interactions mediated by light particles, due to the dependence of the cross section on the momentum transfer and hence difficultly in comparing limits with other bounds~\cite{Alvey:2022pad}.
Since we are interested in how the constraints on the mediator impact the parameter space relevant for cosmic-ray upscattering, 
we shall derive bounds on the product of couplings $g_N g_\chi/ 4\pi$ as a function of mediator mass for a set of benchmark DM masses ($m_\chi =$ 1~MeV, 10~MeV, 100~MeV and 1~GeV). Nevertheless, for comparison with prior work, we also derive bounds on the non-relativistic SI nucleon cross section (i.e. with $q^2\rightarrow 0$) for two benchmark mediator masses. The $1/m_\text{med}^4$ proportionality of the non-relativistic cross section produces absolute values that are significantly larger for the light mediator than heavy case.
\\

We consider data from the LZ, Xenon1T, Borexino and Super-Kamiokande (Super-K) experiments. A summary of the specifications for each of these experiment is given in table~\ref{Tab:Detectors}.
Our bounds are placed by calculating the coupling that gives the 90\% confidence limit on additional events from CRDM. The total number of events is found by integrating Eq.~(\ref{eq:diffrate}) in the range of accessible recoil energies, including the energy-dependent detection efficiency.


\begin{figure}[t]    \includegraphics[width=1.0\columnwidth]{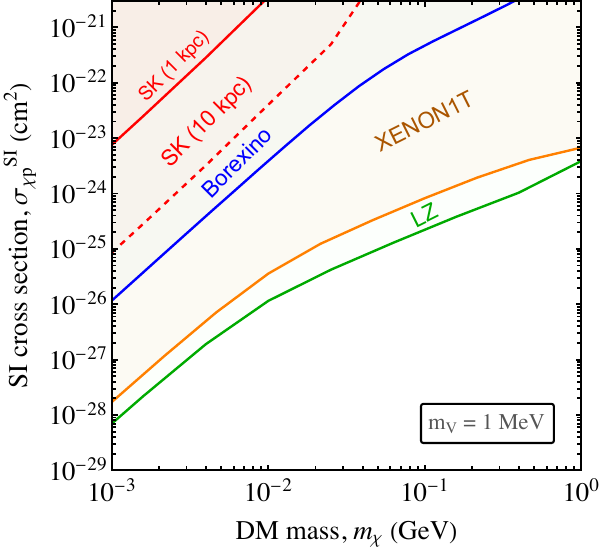}\\
    \includegraphics[width=1.0\columnwidth]{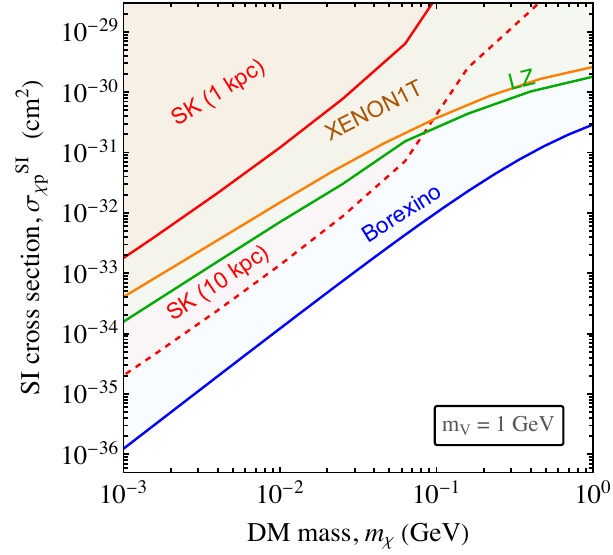}
    \caption{Comparison of CRDM limits on the DM-nucleon cross-section (in the non-relativistic limit) obtained from direct detection and neutrino experiments, for vector mediators of mass $m_v = 1 \,\text{MeV}$\textit{(top)} and $m_v = 1 \,\text{GeV}$ \textit{(bottom)}.}
    \label{fig:limit_comparision}
\end{figure}

\begin{table}[h]
\centering
\caption{The nuclear-recoil energy, live-time and fiducial mass of the experiments used in this analysis.}
\label{Tab:Detectors}
\begin{tabular}{c c c c} 
 \hline \hline
 Experiment & Energy range & Live time & Mass\\ [.5ex] 
 \hline\\
  XENON1T  & 
  [5-60 keV] &
  278.8 days &
  1.3 t \vspace{0.2cm}\\ 
 \hline\\
  Borexino  & 
  [12.3-24.6 MeV] &
  446 days &
  100 t \vspace{0.2cm}\\ 
 \hline\\
 Super-K &
 [0.585-1.54 GeV] &
 6050.3 days &
 22.5 kt \vspace{0.2cm}\\
 \hline \\
 LZ &
 [1-68 keV] &
 60 days &
 5.5 t\\ [1ex]
 \hline \hline
\end{tabular}
\end{table}

The LZ experiment is the largest liquid-xenon time-projection chamber experiment with a fiducial mass of 5.5 tonne. Using preliminary data of only 60 days allowed the LZ collaboration to set the most stringent limits on SI scattering of non-relativistic DM with $m_\chi > 10 $ GeV~\cite{LZ:2022lsv}. For simplicity, we perform a cut-and-count procedure, considering the nearly background-free region below the median nuclear-recoil band (in the S1-S2 space). In this region, only 1 event was observed while 1.4 were expected, corresponding to an upper limit of 2.3 CRDM events. We take the detection efficiency from~\cite{LZ:2022lsv} with an additional 50\% applied to account for the nuclear-recoil cut. While xenon targets offer some SD sensitivity, we will not make use of it here as Borexino proves to be a much more powerful probe.

For comparison with prior work we also compute SI limits using data from the 1 tonne-year exposure of the Xenon1T experiment \cite{XENON:2018voc}. Given the observation of 14 events with an expected background of 7.36, we calculate the coupling that would produce an upper limits of 12 CRDM events. An additional 50\% is also applied to the detection efficiency from \cite{XENON:2018voc}.\\ 

The neutrino experiments Borexino and SuperK offer interesting complementarity, and some advantages over, the xenon-based direct detection experiments. Specifically, the xenon experiments have very low energy thresholds but small target mass, while the neutrino experiments have higher thresholds but significantly larger target mass. This large target mass provides an important advantage, even in the case of SI scattering, where the xenon experiments benefit from an SI cross section enhancement proportion to the square of the atomic number.

For Borexino, following the analysis in \cite{PhysRevD.88.072010} we consider events in the electron-equivalent range of $E_e = 4.8$ - 12.8 MeV (corresponding to proton recoil energies in the range $E_R=12.3$ - 24.6 MeV). We ignore scattering on carbon since it is kinematically suppressed and significantly quenched~\cite{Yoshida2010-tw}, which makes its contribution to the signal rate marginal for the region-of-interest. When considering only hydrogen as a target, there is no nuclear cross section enhancement and thus the experiment is equally sensitive to SI and SD scattering. In the absence of detailed specifications we have assumed a constant detection efficiency of 1, noting that the uncertainties associated with this choice are likely small in comparison to those arising from other simplifications. 

Finally, we use Super-K data following the analysis of \cite{Super-Kamiokande:2022ncz}. Ref.~\cite{Super-Kamiokande:2022ncz} searched for an excess of proton recoil events from the direction of galactic centre, assuming $\rho_\chi=0.42 \,\text{GeV cm}^{-3}$ and a diffusion region of 10~kpc when calculating the DM flux \cite{Ema:2020ulo}. The latter is required to encompass the galactic center and thus provide a directional signal, but incurs significant dependence on the assumed cosmic-ray spectrum and DM density profile near the galactic center. For consistency with other bounds we use $\rho_\chi=0.3 \,\text{GeV cm}^{-3}$ and show the effect of the larger diffusion region. We assume that all protons in SK contribute to the scattering rate, ignoring nuclear binding effects (a valid approximation at the recoil energies considered).

To derive exclusion limits at 90\% confidence we use the log-likelihood ratio statistic,
\begin{equation}
    q_{\mu} = \begin{cases}
                -2 \mathrm{ln}\frac{\mathcal{L}(\mu\vert \hat{\theta})}{\mathcal{L}(\hat\mu\vert \hat{\hat{\theta}})} & \mu \geq \hat\mu\\
                0 & \mu < \hat\mu\\
                \end{cases}
\end{equation}
where the likelihood, $\mathcal{L}(\mu,\theta)$, is taken to be Poissonian and is a function of the signal strength parameter, $\mu$, and nuisance parameters, $\theta$. The hatted parameters indicate maximizing the likelihood with respect to that parameter. Where available, we have profiled over uncertainties in the background. All experiments are taken to have a single bin, with the exception of Borexino where we consider 34 bins in the region-of-interest.
 
Figure~\ref{fig:limit_comparision} shows the bounds on the SI nucleon cross section, for a light (1 MeV) and heavy (1 GeV) vector mediators. The relative strength of these bounds can be understood in terms of the experimental thresholds shown in Fig.~\ref{fig:CRDM Flux}. 
For sub-GeV DM with light mediators, the combination of a low threshold and a large coherent enhancement allows LZ to set the strongest constraints on the SI scattering cross section. 
In the heavy mediator case ($m_\text{med}\gtrsim 200\,\text{MeV}$), the low-energy flux enhancement exhibited for light mediators is no longer relevant. This enables the (larger exposure) neutrino experiments to  become competitive with the (lower-threshold) direct detection experiments. The strongest constraints are then set by scintillator-based Borexino detector, which has a lower threshold (12.3 MeV) than the water Cherenkov-based SK detector (585 MeV).

\begin{figure*}
    \includegraphics[width=1.0\columnwidth]{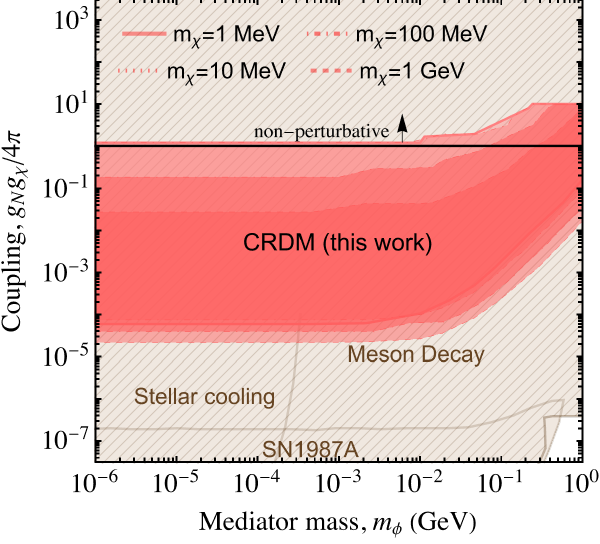}
  \hfill
  \includegraphics[width=1.0\columnwidth]{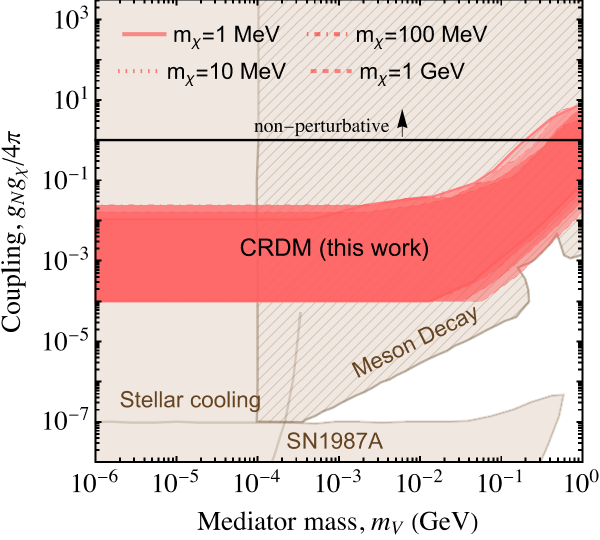}
  \caption{SI results:
Constraints on the sub-GeV scalar \textit{(left panel)} and vector \textit{(right panel)} mediator coupling to the nucleon, from the direct detection of CRDM in Xenon1T, Borexino, SK and LZ, for DM masses $m_\chi=$ 1 MeV, 10 MeV, 100 MeV, 1 GeV. For comparison, we show the regions where the new light-mediator is ruled out by prior constraints.}
\label{fig:Results_SI}
\end{figure*}

\begin{figure*}[t]   
  \includegraphics[width=1.0\columnwidth]{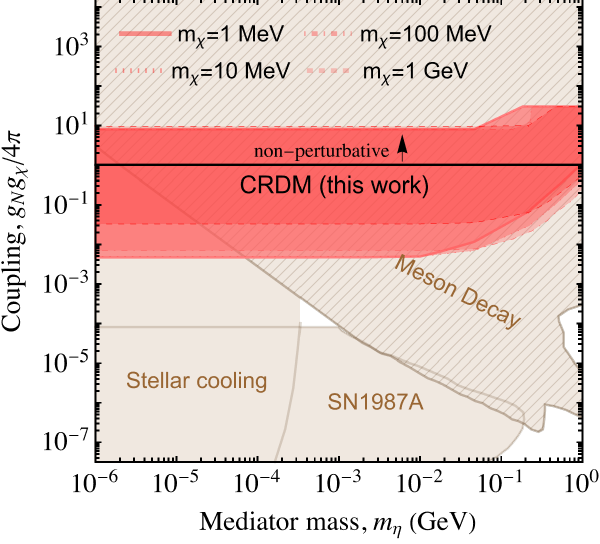}
  \hfill   
  \includegraphics[width=1.0\columnwidth]{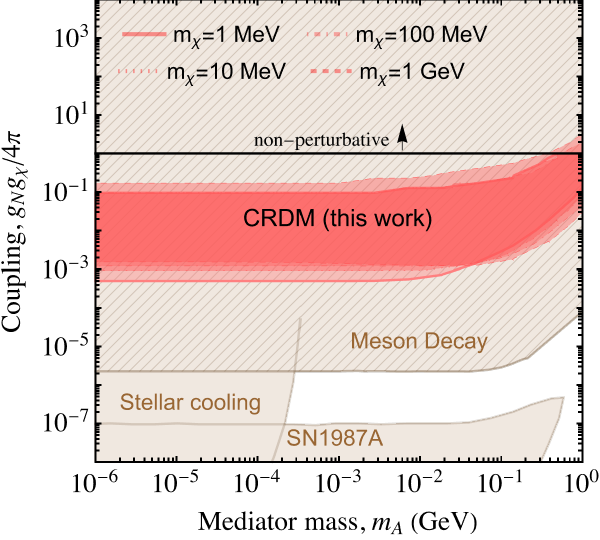}
  \caption{SD results:
 Constraints on the sub-GeV pseudoscalar \textit{(left panel)} and axial-vector \textit{(right panel)} mediator coupling to the nucleon, from the direct detection of CRDM in Borexino and SK, for DM masses $m_\chi=$ 1 MeV, 10 MeV, 100 MeV, 1 GeV. For comparison, we show the regions where the new light-mediator is ruled out by prior constraints.}
  \label{fig:Results_SD}
\end{figure*}

As discussed above, expressing the bounds in terms of cross section are problematic, but it also makes it difficult to compare the limits with direct constraints on the light mediator. Therefore, we now consider constraints on the couplings and mediator masses for all four mediator models introduced in section~\ref{sec:models}.  The scalar and vector mediators will interact via SI scattering while the pseudoscalar and axial vector will interact via SD scattering. We determine bounds on the SI interactions using all the experiments discussed above, while the SD interactions are better probed by those with hydrogen targets. We then display the union of the resulting bounds for a series of DM masses. The SI bounds are shown in Fig.~\ref{fig:Results_SI} while the SD bounds are shown in Fig.~\ref{fig:Results_SD}. We compare our bounds to those derived from (model-dependent) rare meson decays \cite{Knapen:2017xzo} (scalar), \cite{Dror:2017ehi} (vector), 
\cite{Dolan:2014ska} (pseudoscalar), \cite{Kahn:2016vjr} (axial vector), stellar cooling limits from HB and RG stars \cite{Hardy:2016kme}, and SN1987A trapping and cooling limits \cite{Rrapaj:2015wgs}. 

The meson decay constraints are model dependent, since the effective nucleon couplings in Eqs.~(\ref{eq:modelS})-(\ref{eq:modelPS}) are sensitive to choice of quark couplings. The prescription for mapping quark couplings to nucleon couplings is given in e.g.~\cite{Ema:2020ulo,Dent:2015zpa,Del_Nobile2021-dx}. In displaying the meson decay constraints on the nucleon coupling, we have adhered to the minimal flavor violation hypothesis. In practice this means we have assumed the scalar and pseudoscalar couple to all the quarks proportional to their masses, i.e. $g_q \propto m_q/v$. This has a particularly large effect on the scalar mediator case, where decays via top-quark loops dominate the bounds. While the model-dependence of meson decay searches imply that it may, in principle, be possible to weaken these constraints, models that could achieve this would likely be difficult to UV complete. Some model building attempts have been sketched in Refs~\cite{PhysRevLett.130.031803,Ema:2020ulo,Knapen:2017xzo}, which introduced new heavy degrees of freedom that couple only to a subset of the quark generations; these models are still highly constrained.
Lastly, when displaying the meson constraints on $g_N$ in the space of $g_N g_\chi/4\pi$, we conservatively assume $g_\chi=4\pi$. Therefore any reasonable choice of $g_\chi<4\pi$ weakens the CRDM constraints with respect to the meson bounds.

Stellar cooling limits on the scalar boson mediator \cite{Hardy:2016kme} are adapted to the vector by making it a factor of two more restrictive, as the two polarization states result in an increased emission rate \cite{Raffelt:1996wa}. Similarly, the SN1987A bounds from Ref. \cite{Rrapaj:2015wgs} on the light vector boson are made a factor of two less restrictive when applied to the scalar. Note that we adopt the vector constraints on the axial vector, however the axial coupling is likely to have stronger astrophysical constraints than those on a purely vector coupling. 

All of the CRDM bounds show similar characteristics: the constraints are mediator mass independent until they weaken at approximately $m_\text{med}\sim 10$ MeV, where $m_\text{med}^2 \ll q^2$ no longer holds. In all cases, except the vector case, the region where attenuation becomes important is in the non-perturbative coupling regime. Note the difference in the upper bounds of the vector and scalar models. This difference can be traced back to the different dependence of the cross sections on the incident kinetic energy $T_i$.

The results in Figs.~\ref{fig:Results_SI}, show that, for all models considered, there is no additional parameter space where constraints from CRDM using current experiments surpass the existing constraints on the mediators themselves. 

Future detectors would require a significant improvement in sensitivity to be able to probe unconstrained regions for parameter space for these models - i.e. the region below the SN1987A bounds (scalar and pseudoscalar cases) or below the meson bounds (vector and axial vector cases). Furthermore, it should be noted that in this parameter space the limits are improved proportional to the fourth-root of the exposure, and thus large increases in exposure are required for modest gains in sensitivity.

\section{Conclusions}
Using a general set of simplified models for DM-nucleon interactions, we have shown that constraints on the mediating particle greatly restrict the viability of CRDM as a probe of the DM parameter space.  While CR upscattering of DM may allow for the exploration of sub-GeV DM with existing DM direct detection and neutrino experiments, our results highlight the model dependence of CRDM studies and the importance of considering constraints on the particle mediating the DM-nucleon interactions.

We derived constraints on the mediator couplings arising from elastic scattering of CRDM in the XENON1T, LZ, Borexino and Super-K detectors. The low threshold and nuclear cross section enhancement of the xenon targets in LZ produced the strongest bounds across the parameter space for the SI models. For SD models, Borexino produced the strongest bound.  For all simplified models considered, we find that there is essentially no regime where the direct detection of CRDM is not already ruled out by constraints on the mediator. 

While future detectors such as JUNO, DUNE and Hyper-K will improve the CRDM limit (albeit at a slow pace), the existing constraints on the mediators already exclude large regions of the parameter space within reach of these future experiments. 

Finally, these conclusions are applicable to DM-nucleon couplings. Models in which DM couples to leptons, exclusively or in addition to hadrons, would be interesting to study. 
\\
\begin{acknowledgments}
We thank Linyan Wan for her detailed discussions about the Super-K results. The work of NFB, JLN \& ISA is supported by the Australian Research Council through the ARC Centre of Excellence for Dark Matter Particle Physics, CE200100008. ISA is supported by a University of Melbourne Graduate Research Scholarship.
\end{acknowledgments}

\bibliography{Bib.bib}

\end{document}